\documentclass[aps,prb,twocolumn,showpacs,preprintnumbers,amsmath,amssymb,superscriptaddress]{revtex4}%

\usepackage{graphicx}%
\usepackage{dcolumn}
\usepackage{amsmath}

\begin{document}

\preprint{PREPRINT (\today)}

\title{Study of the in-plane magnetic penetration depth
in the cuprate superconductor Ca$_{2-x}$Na$_x$CuO$_2$Cl$_2$: role
of the apical sites}

\author{R.~Khasanov}
 \affiliation{Physik-Institut der Universit\"{a}t Z\"{u}rich,
Winterthurerstrasse 190, CH-8057 Z\"urich, Switzerland}
\author{N.D.~Zhigadlo}
 \affiliation{Laboratory for Solid State Physics, ETH Z\"urich, CH-8093 Z\"urich,
Switzerland}
\author{J.~Karpinski}
 \affiliation{Laboratory for Solid State Physics, ETH Z\"urich, CH-8093 Z\"urich,
Switzerland}
\author{H.~Keller}
 \affiliation{Physik-Institut der
Universit\"{a}t Z\"{u}rich, Winterthurerstrasse 190, CH-8057
Z\"urich, Switzerland}

 \begin{abstract}
A study of the in-plane magnetic penetration depth $\lambda_{ab}$ in
a series of the cuprate superconductors
Ca$_{2-x}$Na$_x$CuO$_2$Cl$_2$ (Na-CCOC) with Na content
$x\simeq0.11$, 0.12, 0.15, 0.18, and 0.19 is reported. The zero
temperature values of $\lambda_{ab}(0)$ were obtained by means of
the muon-spin rotation technique, as well as from measurements of
the intrinsic susceptibility $\chi^{int}(0)$ by using the procedure
developed by Kanigel {\it et al.} [Phys.~Rev.~B {\bf 71}, 224511
(2005)]. $\lambda_{ab}$ at $T=0$~K was found to increase with
decreasing doping from $\lambda_{ab}(0)=316(19)$~nm for the
$x\simeq0.19$ sample to $\lambda_{ab}(0)=430(26)$~nm for the
$x\simeq0.11$ one. From a comparison of the present Na-CCOC data
with those of Bi2201 and La214 cuprate superconductors it is
concluded that substitution of the apical oxygen by chlorine
decreases the coupling between the superconducting CuO$_2$ planes,
leading to an enhancement of the two-dimensional properties of
Na-CCOC.
 \end{abstract}

\maketitle

\section{Introduction}

The question if the out-of-plane electronic states are essential for
high-temperature superconductivity is a matter of debate for a long
time. In fact, the idea that in perovskites with their octahedral
oxygen environment, the Jahn-Teller effect may lead to an enhanced
electron-phonon coupling has guided Bendorz and M\"uller in their
original search for oxide superconductors.\cite{Bednorz86} Bearing
this in mind one may ask if, in addition to the planar states, the
apical states too play a role in the occurrence of high-temperature
superconductivity? There is still no clear answer to this question.
On the one hand, observation of superconductivity in materials with
apical oxygen replaced by halogen atoms (like F, Cl, Br),
\cite{Al-Mamouri94,Hiroi94,Zenitani05} as well as the absence of an
apical oxygen isotope effect on the transition temperature
$T_c$\cite{Zech94} and the zero-temperature in-plane magnetic field
penetration depth $\lambda_{ab}(0)$\cite{Khasanov03} suggest that
the apical sites do not play a major role. On the other hand, the
doubling of $T_c$\cite{Locquet98} and the change of the Fermi
surface topology from ''hole-like`` to
''electron-like``\cite{Pavuna03} in epitaxially strained
La$_{2-x}$Sr$_x$CuO$_4$ films, as well as a clear correlation
between the transition temperature at optimal doping and the degree
of localization of the axial orbitals in the CuO$_2$
planes\cite{Pavarini01} give evidence that apical states are indeed
involved in superconductivity.  One should also mention that there
are compounds where superconductivity has proven to be induced by
apical oxygen doping.\cite{Jin95,Yang07}

In order to elucidate the role of the apical sites, it is important
to clarify the origin of similarities and differences between
cuprates with oxygen and halogen atoms (like F, Cl, Br ) on the
apical sites. Crucial information can be obtained from measurements
of the magnetic field penetration depth $\lambda$. $\lambda$ is one
of the fundamental lengths of a superconductor which, within a
simple London model, relates two important superconducting
parameters: the charge carrier concentration $n_s$ and the mass of
the charge carriers $m^\ast$, according to $\lambda^{-2}\propto
n_s/m^\ast$. The temperature dependence of $\lambda$ reflects the
quasiparticle density of states available for thermal excitations
and, therefore, probes the superconducting gap structure. The shape
of $\lambda(T)$ and the zero-temperature value $\lambda(0)$ provides
relevant information on the superconducting mechanism and sets a
scale for the screening of an external magnetic field. Here we
report studies of the in-plane magnetic penetration depth
$\lambda_{ab}$ for a series of Ca$_{2-x}$Na$_x$CuO$_2$Cl$_2$
(Na-CCOC) samples with $x\simeq0.11$, 0.12, 0.15, 0.18, and 0.19.
Ca$_{2-x}$Na$_x$CuO$_2$Cl$_2$ is a structural analogue to
La$_{2-x}$Sr$_x$CuO$_4$ with Cl atoms replacing oxygen on the apical
sites. The zero temperature values of $\lambda_{ab}$ were obtained
by means of muon-spin rotation ($\mu$SR), as well as from
measurements of the intrinsic susceptibility $\chi^{int}(0)$ by
using the procedure developed by Kanigel {\it et
al.}\cite{Kanigel05} It was found that the measured
$\lambda_{ab}^{-2}(0)$ data points for Na-CCOC are all shifted to
the left side of the universal ''Uemura`` line, that represents a
linear correlation between the zero-temperature superfluid density
$\rho_s\propto\lambda_{ab}^{-2}(0)$ and the transition temperature
$T_c$ for various hole-doped high-temperature cuprate
superconductors (HTS's).\cite{Uemura89,Uemura91} Based on a
comparison of the Na-CCOC data with the data of
Bi$_2$Sr$_{2-x}$La$_x$CuO$_{6+\delta}$ (Bi2201) and
La$_{2-x}$Sr$_x$CuO$_4$ (La214) it is concluded that replacing of
apical oxygen by Cl  decreases the coupling between the
superconducting CuO$_2$ planes, leading to an enhancement of the
two-dimensional properties of Na-CCOC. The reason for this is very
likely due to a substantial reduction of the amount of holes on the
apical sites in comparison to La214. In addition,  field-induced
magnetism was detected in optimally doped
Ca$_{1.82}$Na$_{0.18}$CuO$_2$Cl$_2$, suggesting that Na-CCOC has a
competing magnetic state very close in free energy to their
superconducting state.

The paper is organized as follows. In Sec.~\ref{sec:experimental}
we describe the sample preparation procedure and details of the
muon-spin rotation and magnetization experiments. The dependence
of the muon-spin depolarization rate $\sigma$ on temperature and
magnetic field is presented in Sec.~\ref{sec:field-effects}.
Secs.~\ref{sec:lambda_muSR} and \ref{sec:lambda_magnetization}
comprise studies of the in-plane magnetic penetration depth
$\lambda_{ab}$ by means of $\mu$SR and magnetization techniques.
The comparison of the superfluid density
$\rho_s\propto\lambda_{ab}^{-2}(0)$ for Na-CCOC with that for
other hole-doped HTS's with oxygen on the apical site is presented
in Sec.~\ref{sec:discussions_3D-superfluid}. Two-dimensional
aspects of the superfluid density are discussed in
Sec.~\ref{sec:discussions_2D-superfluid}. The conclusions follow
in Section~\ref{sec:conclusion}.

\section{Experimental details}\label{sec:experimental}


Underdoped and optimally doped superconducting
Ca$_{2-x}$Na$_x$CuO$_2$Cl$_2$ samples with Na content $x\simeq0.11$,
0.12, 0.15, 0.18, and 0.19 were synthesized under high pressure by
using the procedure described in Ref.~\onlinecite{Zhigadlo07}. It
includes, first, the synthesis of the non superconducting parent
compound Ca$_2$CuO$_2$Cl$_2$, and second, the high pressure
annealing of Ca$_2$CuO$_2$Cl$_2$ mixed with NaClO$_4$ and NaCl.
Ca$_2$CuO$_2$Cl$_2$ was synthesized by a solid-state reaction of
Ca$_2$CuO$_3$, CuO, and CaCl$_2$. The powder mixture was pressed
into a pellet and synthesized at 750~$^\circ$C in argon flow with
several intermediate grindings under ambient pressure. The resulting
Ca$_2$CuO$_2$Cl$_2$ was well mixed with NaClO$_4$ and NaCl in a
molar ratio of 1:0.2:0.2 in a dry box and sealed in Pt cylindrical
capsules of 6-8~mm internal diameter and 7-9~mm length.
High-pressure experiments were performed in  opposed anvil-type
high-pressure devices at 40-45~kbar. After applying pressure the
temperature was increased during 1.5~h up to the maximum of
1350-1700~$^\circ$C and kept stable for 0.5~h. Then the temperature
was slowly decreased to 1000~$^\circ$C and, finally, to room
temperature. The high pressure was maintained constant throughout
the synthesis and was removed only after the cell was cooled to room
temperature.
The Na content was estimated from the comparison of the $c-$axis
lattice constants obtained in x-ray experiments with those
reported in Refs.~\onlinecite{Hiroi94} and \onlinecite{Kohsaka02}.
Due to the extreme high hygroscopicity of
Ca$_{2-x}$Na$_x$CuO$_2$Cl$_2$, all the manipulations with the
samples were made in a glove box filled with Ar.

\begin{figure}[htb]
\includegraphics[width=0.8\linewidth]{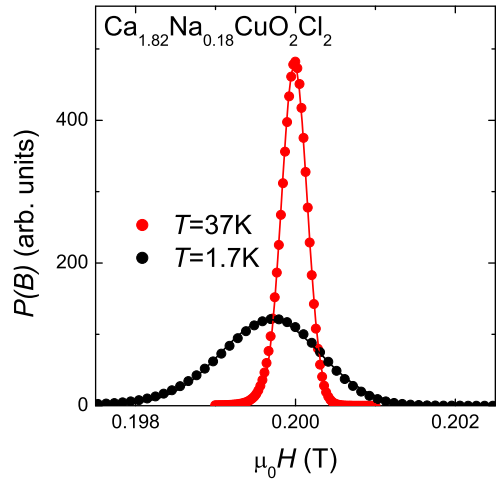}
%
\caption{(Color online) Internal magnetic field distribution $P(B)$
for Ca$_{1.82}$Na$_{0.18}$CuO$_2$Cl$_2$ sample at $\mu_0H=0.2$~T in
the normal ($T=37$~K) and the superconducting ($T=1.7$~K) state
obtained by means of the maximum entropy Fourier transform
technique. The solid lines represent fits with a single Gaussian
line. }
 \label{fig:signals}
\end{figure}

Zero-field and transverse-field $\mu$SR experiments were performed
at the $\pi$M3 and $\pi$E1 beam lines at the Paul Scherrer
Institute (Villigen, Switzerland). The $\mu$SR experiments were
performed on two samples: an optimally doped sample with Na
content $x\simeq0.18$ ($T_c\simeq27$~K) and a slightly underdoped
one with $x\simeq0.12$ ($T_c\simeq18$~K).
In a superconducting sample the magnetic penetration depth
$\lambda$ can be extracted from the second moment $\langle \Delta
B^{2}\rangle$ of the probability distribution of the local
magnetic field $P(B)$ in the mixed state probed by
$\mu$SR.\cite{Zimmermann95}
In the present study we first analyzed the $\mu$SR time spectra by
using a direct Fourier transform based on a maximum entropy
algorithm,\cite{Rainford94} with no prior assumptions on the form of
$P(B)$. It was found that a Gaussian distribution of local fields
gives a reasonable estimate of $P(B)$ (see Fig.~\ref{fig:signals}),
in agreement with previous observations. \cite{Pumpin90} Therefore,
the $\mu$SR time spectra were analyzed by using a Gaussian
relaxation function $R(t) = \exp[-\sigma^{2}t^{2}/2]$. The second
moment of $P(B)$ was then obtained as:  $\langle \Delta
B^{2}\rangle=\sigma^2/\gamma_\mu^2$ ($\gamma_{\mu} =
2\pi\times135.5342$~MHz/T is the muon gyromagnetic ratio).
%


The field-cooled 0.5~mT magnetization measurements for the Na-CCOC
samples ($x\simeq0.11$, 0.12, 0.15, 0.18, and 0.19) were performed
by using a SQUID magnetometer. In order to follow the procedure
described in Ref.~\onlinecite{Kanigel05}, the samples used for the
magnetization experiments were preliminary powderized and then
placed in quartz ampoules (1.5~mm inner diameter). The mass of the
samples were $\sim25-30$~mg. The ratio of the diameter to the
height of the powder in the ampoule was approximately 1 to 7. The
magnetic field was applied parallel to the ampoule axis. As shown
in Ref.~\onlinecite{Kanigel05}, for this geometry the zero
temperature ''intrinsic`` susceptibility, obtained from the
field-cooled magnetization, is proportional to the inverse squared
in-plane magnetic penetration depth
$\chi^{int}(0)\propto\lambda_{ab}^{-2}(0)$.

\section{Experimental results}\label{sec:experimental_data}

\subsection{Dependence of the muon-spin depolarization rate $\sigma$ on $T$
and $H$} \label{sec:field-effects}

\begin{figure}[htb]
\includegraphics[width=0.85\linewidth]{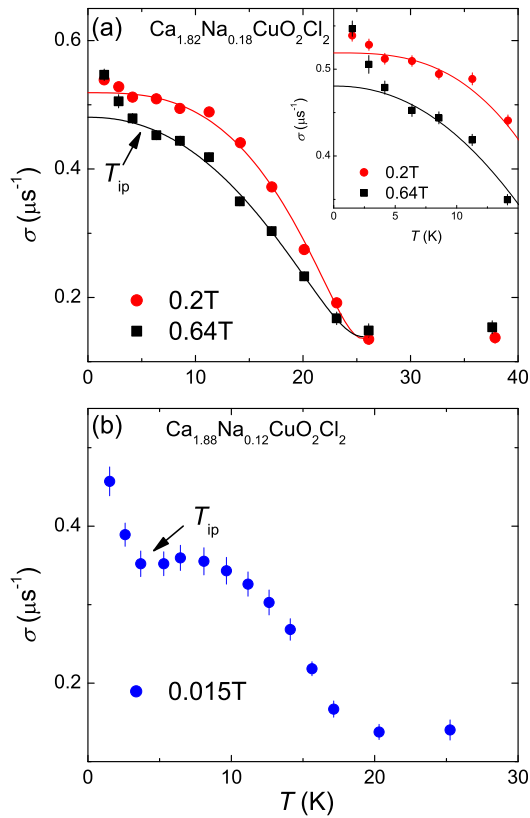}
%
\caption{(Color online) (a) Temperature dependence of the muon-spin
depolarization rate $\sigma$ for Ca$_{1.82}$Na$_{0.18}$CuO$_2$Cl$_2$
measured after field cooling in $\mu_0H=0.2$~T (red circles) and
0.64~T (black squares). (b) $\sigma(T)$ for
Ca$_{1.88}$Na$_{0.12}$CuO$_2$Cl$_2$ measured in $\mu_0H=0.015$~T.
The solid lines in (a) represent fits of Eq.(\ref{eq:power-law}) to
the data. See text for details. The inset in (a) shows the extension
of the low-temperature region. }
 \label{fig:T-scan}
\end{figure}

Figure~\ref{fig:T-scan} shows the temperature dependences of the
muon-spin depolarization rate $\sigma$ obtained from the fits to
the $\mu$SR data.  It is seen that for both Na-CCOC samples
studied by $\mu$SR, $\sigma(T)$ is constant above $T_c$ and starts
to rise with decreasing temperature for $T<T_c$. Most
interestingly, however, is that in the low-temperature region an
inflection point ($T_{ip}$) appears below which $\sigma$ starts to
increase rather sharply. It is also seen that with increasing
magnetic field, $\sigma$ below $T_{ip}$ rises faster and $T_{ip}$
has a tendency to shift to higher temperatures. Indeed, the inset
in Fig.~\ref{fig:T-scan}~(a) shows that when the magnetic field
increases from 0.2~T to 0.64~T, $T_{ip}$ for the $x\simeq0.18$
sample shifts from approximately 4~K to 6~K, and the ratio
$\sigma(1.7$~K)/$\sigma(T_{ip})$ changes from 1.05 to 1.20.
The behavior of $\sigma(T)$ below $T_{ip}$ clearly demonstrates
that some kind of magnetic ordering takes place in Na-CCOC. The
zero-field $\mu$SR experiments of Ohishi {\it et
al.},\cite{Ohishi05} performed on a series of Na-CCOC
($x=0.0\div0.12$) samples, reveal the presence of spin-glass type
magnetism in the $x=0.12$ sample with a spin-glass ordering
temperature of $\simeq2.5$~K. This suggests that the increase of
$\sigma$ at $T<T_{ip}$, seen in Fig.~\ref{fig:T-scan}~(b), is
simply a consequence of it.
The situation for the $x\simeq0.18$ sample is, however, not so
clear. Our zero-field $\mu$SR experiments gave no indication for
{\it any} magnetism in this particular sample down to
$T\simeq1.7$~K. On the other hand, the increase of $\sigma(T)$ below
$T_{ip}$ and the shift of $T_{ip}$ to higher temperatures, both
correlated with the magnetic field [see inset in
Fig.~\ref{fig:T-scan}~(a)], clearly demonstrate the appearance and
{\it enhancement} of magnetism in the $x\simeq0.18$ sample. Note
that field-induced magnetism was recently observed by Savici {\it et
al.}\cite{Savici05} in highly underdoped La$_{2-x}$Sr$_x$CuO$_4$,
La$_{2-x}$Ba$_x$CuO$_4$, and La$_{2-x}$Eu$_x$CuO$_4$ HTS's. It was
shown, that the increase of the relaxation above the
antiferromagnetic ordering temperature $T_N$ and the superconducting
transition temperature $T_c$ is due to quasistatic random fields
induced by the external magnetic field.\cite{Savici05}

\begin{figure}[htb]
\includegraphics[width=0.9\linewidth]{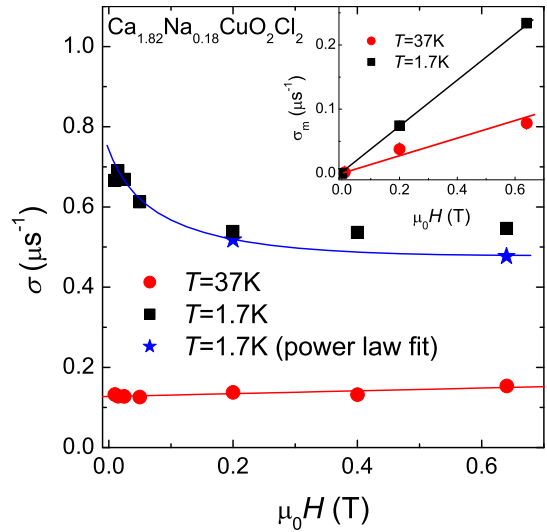}
%
\caption{(Color online) Magnetic field dependence of the muon-spin
depolarization rate $\sigma$ of Ca$_{1.82}$Na$_{0.18}$CuO$_2$Cl$_2$
above (red circles) and below (black squares) the superconducting
transition temperature. Stars are the values of $\sigma(T=1.7$~K)
obtained after substraction of the magnetic contribution $\sigma_m$.
The blue and the red lines are guides to the eye. The inset shows
$\sigma_m(H)$ at $T=37$~K (red circles) and $T=1.7$~K (black
squares). See text for details. }
 \label{fig:field-scan}
\end{figure}

In order to study field-induced magnetism in the $x\simeq0.18$
sample in more detail, $\sigma$ was measured as a function of $H$
above (37~K) and below (1.7~K) the superconducting transition
temperature (see Fig.~\ref{fig:field-scan}). Note that all
$\sigma(1.7$~K) data points were obtained after field-cooling the
sample from far above $T_c$ to 1.7~K in the corresponding field.
Above $T_c$ the magnetic contribution $\sigma_{m}$ was obtained by
subtracting the nuclear contribution
($\sigma_{nm}=0.132$~$\mu$s$^{-1}$ at $\mu_0H=0.01$~T) from the
measured $\sigma$ as $\sigma_{m}^2=\sigma^2-\sigma_{nm}^2$.
In order to obtain $\sigma_{m}$ at $T=1.7$~K we used the following
procedure. Bearing in mind that below $T_c$ a superconducting
component $\sigma_{sc}$ is present, the total $\sigma$ obtained in
$\mu$SR experiments is a sum of:
\begin{equation}
\sigma^2(T)=\sigma^2_{m}(T)+\sigma_{sc}^2(T)+\sigma_{nm}^2(T).
 \label{eq:sigma-tot}
\end{equation}
In a next step we assumed that $\sigma_{nm}$ does not depend on
temperature and that in the temperature range $T_{ip}<T<T_c$ the
measured $\sigma(T)$ is determined predominantly by the
superconducting part of the muon-spin depolarization rate
$\sigma_{sc}(T)$ and the nuclear dipolar  component $\sigma_{nm}$.
The temperature dependence of $\sigma_{sc}(T)$ was assumed to be
described by the power law:\cite{Zimmermann95}
\begin{equation}
\sigma_{sc}(T)=\sigma_{sc}(0)[1-(T/T_c)^n].
 \label{eq:power-law}
\end{equation}
The corresponding fitting curves are shown in
Fig.~\ref{fig:T-scan}~(a). Finally, $\sigma_{m}$ at $T=1.7$~K was
calcualted as $\sigma_m(1.7$~K)$=[\sigma^2(1.7~{\rm
K})-\sigma_{sc}^2(1.7~{\rm K})-\sigma_{nm}^2]^{0.5}$. The field
dependence of $\sigma_{m}$ for $T=1.7$~K and 37~K are shown in the
inset of Fig.~\ref{fig:field-scan}. It is seen that for both
temperatures $\sigma_m$ increases linearly with increasing magnetic
field, in good agreement with the results of Savici {\it et
al.}\cite{Savici05} We want to point, however, to the difference
between the optimally doped Ca$_{1.82}$Na$_{0.18}$CuO$_2$Cl$_2$
sample measured here and the underdoped La$_{2-x}$Sr$_x$CuO$_4$,
La$_{2-x}$Ba$_x$CuO$_4$, and La$_{2-x}$Eu$_x$CuO$_4$ HTS's studied
in Ref.~\onlinecite{Savici05}. Till now, field-induced magnetism was
observed only for systems exhibiting {\it static} magnetism in zero
field.\cite{Savici05} This is not the case for $x=0.18$ sample,
since no magnetism was detected in zero-field $\mu$SR experiments
down to $\simeq$1.7~K. Even though we cannot rule out completely the
appearance of zero-field magnetism at lower temperatures, the
enhancement of magnetism with increasing field would imply that even
optimally doped Na-CCOC has a competing magnetic state very close in
free energy to their superconducting state.

\subsection{Determination of $\lambda_{ab}(0)$ by means of
$\mu$SR} \label{sec:lambda_muSR}

In earlier transverse-field $\mu$SR experiments on
YBa$_2$Cu$_3$O$_{7-\delta}$ it was observed that the relaxation rate
$\sigma$ for nonoriented powders exhibits nearly no magnetic field
dependence in a rather broad field range (typically, $\sim$~0.05~T
to $\sim$~0.4~T).\cite{Zimmermann95} This can be explained within
the London model which predicts that the second moment of the
magnetic field distribution in a perfect vortex lattice is
independent of the applied magnetic field for $2H_{c1} \lesssim H\ll
H_{c2}$ ($H_{c1}$ and $H_{c2}$ are the first and the second critical
fields, respectively).\cite{Fesenko91}
This feature allows a direct comparison of values of $\sigma_{sc}$
obtained for various HTS's at various doping levels and taken at
different magnetic fields.
In order to check if the relation $\sigma_{sc}(H)\simeq const$
also holds for Na-CCOC  we subtract the magnetic contribution
$\sigma_{m}$ for $\mu_0H=0.2$~T and 0.64~T from the measured
$\sigma$ in accordance with Eq.~(\ref{eq:sigma-tot}) and plot the
resulting $(\sigma_{sc}^2+\sigma_{nm}^2)^{0.5}$ values (blue
stars) in Fig.~\ref{fig:field-scan}. Two tendencies are clearly
seen: First, for fields smaller than 0.2~T one can almost neglect
the magnetic contribution $\sigma_m$ to the measured $\sigma$.
This is confirmed by the field dependence of $\sigma_{m}$
presented in the inset of Fig.~\ref{fig:field-scan}, revealing
that for $\mu_0H<0.2$~T $\sigma_m(1.7~K)$ is more than 10 times
smaller than the measured $\sigma$. Second, the solid blue line
shows that for the $x\simeq0.18$ sample the relation
$\sigma_{sc}(H)\simeq const$ holds for $\mu_0H\gtrsim0.15$~T. The
slow decrease of $\sigma_{sc}$ above 0.15~T can be explained by
nonlocal and nonlinear corrections to $\sigma_{sc}$ due to the
$d-$wave order parameter of
HTS.\cite{Sonier00,Amin00,Khasanov07_field-effect}

\begin{figure}[htb]
\includegraphics[width=0.9\linewidth]{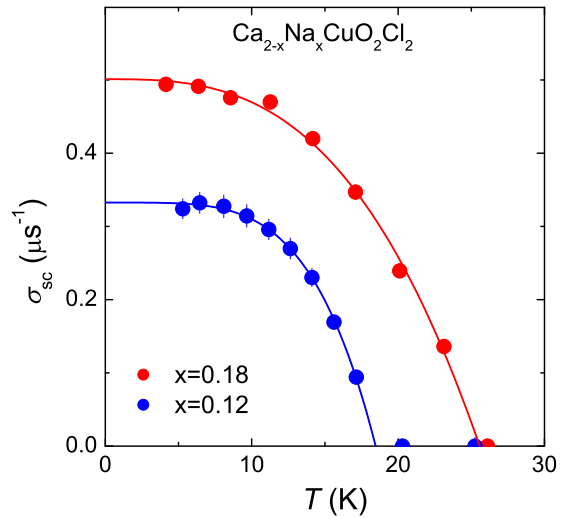}
%
\caption{(Color online) Temperature dependence of
$\sigma_{sc}\propto \lambda_{ab}^{-2}$ for
Ca$_{2-x}$Na$_x$CuO$_2$Cl$_2$ [$x\simeq0.18$ -- upper curve and
$x\simeq0.12$ -- lower curve]. The solid lines represent power law
fits to the data. See text for details. }
 \label{fig:lambda_vs_T}
\end{figure}

Figure~\ref{fig:lambda_vs_T} shows the temperature dependence of
$\sigma_{sc}$, obtained after substraction of $\sigma_{nm}$ from
the data presented in Fig.~\ref{fig:T-scan}. Since at low
temperatures a magnetic contribution is present, data points below
4~K were excluded in the analysis. The solid lines represent fits
to the power law [Eq.~(\ref{eq:power-law})]. From the obtained
values $\sigma_{sc}(0)$ the zero temperature values of the
in-plane magnetic field penetration depth $\lambda_{ab}(0)$ were
determined as:\cite{Zimmermann95}
\begin{equation}
\lambda_{ab}({\rm nm})=\frac{224}{\sqrt{\sigma_{sc}(\mu{\rm
s}^{-1})}}\ .
 \label{eq:lambda_ab}
\end{equation}
The results of the fits and the values of $\lambda_{ab}(0)$
obtained by Eq.~(\ref{eq:lambda_ab}) are summarized in
Table~\ref{Table:lambda_results}. Note that
$\lambda_{ab}(0)=317(19)$~nm for $x\simeq0.18$ sample is more than
40\% smaller than $\lambda_{ab}(0)=438-453$~nm reported in
Ref.~\onlinecite{Kim06}. We do not have any explanation for this
difference, but want to emphasize that $\lambda_{ab}$ directly
derived by $\mu$SR is more reliable than the one obtained from
measurements of the reversible magnetization.\cite{Kim06}

\begin{table}[htb]
\caption[~]{\label{Table:lambda_results} Summary of the $\lambda(T)$
study of Ca$_{2-x}$Na$_x$CuO$_2$Cl$_2$ (see text for details).
} %
\begin{center}
\begin{tabular}{lcccccccc}\\ \hline
\hline
 Method&$x$ &$T_c$&$\sigma_{sc}(0)$&$\chi^{int}(0)$&$\lambda_{ab}(0)$\\
&&(K)&($\mu$s$^{-1}$)&&(nm)\\
\hline

$\mu$SR &0.12 &18.5(2)&0.33(2)& -- & 390(24)\\
        &0.18 &25.6(2)&0.50(3)& -- & 317(19)\\

\hline

              &0.11 &15.14(4)&--&0.171 &430(26)\\
              &0.12 &18.40(3)&--&0.192 &406(24)\\
$\chi^{int}(0)$  &0.15 &23.45(3)&--&0.307&321(19)\\
              &0.18 &27.11(3)&--&0.315&317(19)$^a$\\
              &0.19 &27.70(3)&--&0.316&316(19)\\

 \hline \hline \\

\end{tabular}
   \end{center}
   $^a${\small Normalized to $\lambda_{ab}(0)=317(19)$~nm obtained for
   $x\simeq0.18$
   sample by $\mu$SR}

\end{table}

\subsection{Determination of $\lambda_{ab}(0)$
in low-field magnetization experiments}
\label{sec:lambda_magnetization}

In order to complete the in-plane magnetic field penetration depth
study of the HTS Na-CCOC we performed similar field-cooled
magnetization ($M_{FC}$) experiments as reported by Kanigel {\it et
al.}\cite{Kanigel05} It was shown that for HTS powder samples,
shaped in a cylindrical container having a diameter much smaller
than its length, $\lambda_{ab}(0)$ can be obtained from the
so-called intrinsic susceptibility $\chi^{int}(0)=M_{FC}(0)/M_{id}$
($M_{id}$ is the magnetization of an ideal diamagnet) according to
the relation $\chi^{int}(0)\propto
\lambda^{-2}_{ab}(0)$.\cite{Kanigel05}
\begin{figure}[htb]
\includegraphics[width=0.9\linewidth]{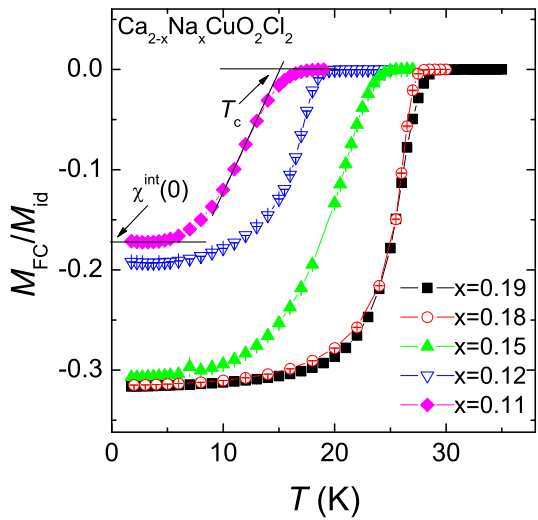}
%
\caption{(Color online) Temperature dependence of the field-cooled
magnetization $M_{FC}$ ($\mu_0H=0.5$~mT) for
Ca$_{2-x}$Na$_x$CuO$_2$Cl$_2$, normalized to the magnetization of an
ideal diamagnet $M_{id}$. From the left to the right $x\simeq0.11$,
0.12, 0.15, 0.18, and 0.19. }
 \label{fig:magnetization}
\end{figure}
Figure~\ref{fig:magnetization} shows $M_{FC}(T)$ curves for
Ca$_{2-x}$Na$_x$CuO$_2$Cl$_2$ ($x\simeq0.11$, 0.12, 0.15, 0.18, and
0.19) samples taken at 0.5~mT. The transition temperatures $T_c$ and
the values of $\chi^{int}(0)$ were obtained from the intersection of
the linearly extrapolated $M_{FC}$ curves with the $M=0$ line and by
extrapolating the low temperature part of $M_{FC}(T)$ to $T=0$,
respectively (see Fig.~\ref{fig:magnetization} and
Table~\ref{Table:lambda_results}). From the measured values of
$\chi^{int}(0)$ the zero temperature values of $\lambda_{ab}(0)$
were then obtained by normalizing to $\lambda_{ab}(0)$=317(19)~nm
derived for the Ca$_{1.82}$Na$_{0.18}$CuO$_2$Cl$_2$ sample by
$\mu$SR (see Table.~\ref{Table:lambda_results}).
A quick glance at Table~\ref{Table:lambda_results} reveals that the
procedure described in Ref.~\onlinecite{Kanigel05} is indeed
reliable, since the values of $\lambda_{ab}(0)$ for the
$x\simeq0.12$ sample obtained by means of both techniques used in
the present study agree rather well.

\section{Discussions}\label{sec:discussions}

\subsection{Comparison with superfluid densities of HTS's
with oxygen on the apical site}
\label{sec:discussions_3D-superfluid}

\begin{figure}[htb]
\includegraphics[width=0.9\linewidth]{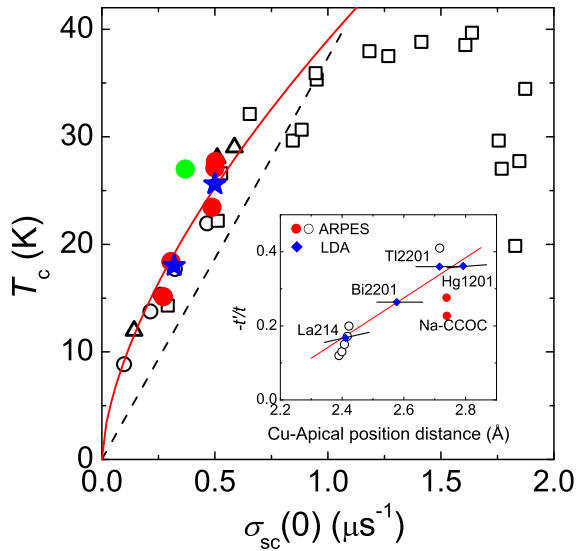}
%
\caption{(Color online) $T_c$ vs. $\sigma_{sc}(0)\propto
\lambda_{ab}^{-2}$ for Na-CCOC and various hole-doped HTS's with
oxygen on the apical site. Solid red circles and blue stars are
Na-CCOC data obtained in the present study (see
Table~\ref{Table:lambda_results}). The green circle is the data
point for Na-CCOC ($x=0.18$) from Ref.~\onlinecite{Kim06}. Open
squares are La214 data from Refs.~\onlinecite{Uemura91} and
\onlinecite{Tallon03}. Open triangles are Bi2201 data taken from
Ref.~\onlinecite{Russo07} and open circles are Y123 data from
Ref.~\onlinecite{Liang05}. The dashed line represents the ''Uemura``
relation.\cite{Uemura89,Uemura91} The solid red line corresponds to
the power law with $\sigma_{sc}(0)\propto T_c^{1.6}$ from
Ref.~\onlinecite{Liang05}. The inset shows the hopping integral
ratio $t'/t$ as a function of plane Cu -- apical position distance
for La214, Bi2201, Tl2201, and Hg2201 HTS's obtained from
local-density approximation (LDA) band structure calculations (after
Ref.~\onlinecite{Pavarini01}). The circles are $t'/t$ values for
Na-CCOC ($x=0.10$, 0.12),\cite{Shen_Thesis} La214,\cite{Yoshida07}
and Tl2201 \cite{Plate05} obtained from ARPES data. See text for
details.}
 \label{fig:uemura}
\end{figure}

To compare the results for Na-CCOC with the results for other
hole-doped superconductors, in Fig.~\ref{fig:uemura} we plot $T_c$
as a function of the zero temperature superfluid density
$\rho_s\propto\lambda^{-2}_{ab}(0)\propto\sigma_{sc}(0)$. We also
include data points for Na-CCOC ($x=0.18$),\cite{Kim06} for the
structurally related compound with oxygen on the apical site
La$_{2-x}$Sr$_x$Cu$_4$ (La214),\cite{Uemura91,Tallon03} for the
single layer HTS Bi$_2$Sr$_{2-x}$La$_x$CuO$_{6+\delta}$
(Bi2201),\cite{Russo07} and for underdoped
YBa$_2$Cu$_3$O$_{7-\delta}$ (Y123) with highly reduced
$T_c$.\cite{Liang05} The dashed line represents the famous
''Uemura`` relation (linear correlation between $T_c$ and
$\lambda^{-2}_{ab}(0)\propto n_s/m^\ast$ for various families of
underdoped HTS's\cite{Uemura89,Uemura91}). It is seen that points
for Na-CCOC at all levels of doping, as well as the points for
Bi2201 and highly underdoped Y123 lie significantly higher than
expected for the ''Uemura`` relation. The solid line corresponds to
the power law $\sigma_{sc}(0)\propto T_c^{1.6}$ obtained in
Ref.~\onlinecite{Liang05}. While the agreement between Na-CCOC,
Bi2201, and highly underdoped Y123 is rather good, the points for
the structurally related compound La214 are shifted to the right.
Only 4 out of 18 points coincide with those for Na-CCOC: three
points for underdoped La214 with highly reduced $T_c$'s and one for
overdoped La214 ($T_c\simeq 32$~K). As was recently pointed out by
Russo {\it et al.}, \cite{Russo07} the agreement with underdoped
Y123 and with points for underdoped La214 should be taken with
caution. It was shown by zero-field $\mu$SR experiments on La214
\cite{Savici02} and Y123 \cite{Sanna04} having rather reduced
$T_c$'s, that a major volume fraction exhibits static magnetic order
and, probably, does not carry the superfluid.\cite{Kojima03} Thus
the reduction of the superfluid density for both of these compounds
may be a simple consequence of it.\cite{Russo07} In contrast, the
results for Bi2201 were obtained for samples which do not involve
static magnetic order \cite{Russo07} and, correspondingly, might
represent an intrinsic property free of possible complications due
to magnetic fractions. Therefore, we are first going to compare the
present Na-CCOC data with those of Bi2201 and, later on, with La214.

The good agreement between the Na-CCOC and Bi2201 data presented
in Fig.~\ref{fig:uemura} suggests that there are some similarities
between these two compounds:
%
(i) Both Na-CCOC and Bi2201 are highly anisotropic
superconductors. Highly two-dimensional (2D) properties of the
Na-CCOC system were recently reported  by Kim {\it et
al.}\cite{Kim06} They found that the fluctuation induced
magnetization and the irreversibility line obtained for Na-CCOC
($x=0.18$) show pronounced 2D behavior. The anisotropy coefficient
$\gamma$ was estimated to be in the range $50<\gamma<800$. Even
though the range of $\gamma$ reported in Ref.~\onlinecite{Kim06}
is rather broad, the value of $\gamma$ is much higher than, {\it
e.g.} $\gamma=15$ obtained for optimally doped
La214,\cite{Panagopoulos00} but is consistent with
$\gamma\simeq400$ for optimally doped Bi2201.\cite{Kawamata99}
(ii) The inset in Fig.~\ref{fig:uemura} reveals that, both Na-CCOC
and Bi2201 have similar values for the hopping integral ratio $t'/t$
($t$ and $t'$ are the first and the second nearest neighbor transfer
integrals between the Cu sites in CuO$_2$ planes). Pavarini {\it et
al.}\cite{Pavarini01} showed that the ratio $t'/t$ is the essential
material-dependent parameter which is mainly controlled by the
energy of the apical orbital. It was also shown that for hole-doped
HTS's the maximum transition temperature for a particular
superconducting family
increases with increasing $t'/t$. In the inset of
Fig.~\ref{fig:uemura} we reproduce the original figure from
Ref.~\onlinecite{Pavarini01}, where $t'/t$ obtained from
local-density approximation (LDA) band structure calculations is
plotted as a function of the plane Cu -- apical position distance
for the single-layer HTS's La214, Bi2201, Tl2201, and Hg2201. In
this figure we also include the values of $t'/t$ for Na-CCOC
($x=0.10$, 0.12),\cite{Shen_Thesis} La214,\cite{Yoshida07} and
Tl2201 \cite{Plate05} obtained from the analysis of Angle Resolved
Photoemission (ARPES) data. It is seen that the $t'/t$ ratio for
Na-CCOC  is very close to the one of Bi2201.

In the next step we compare the Na-CCOC data with those for the
structurally related compound La214. The general difference between
them is that Cl atoms, instead of oxygen atoms, occupy the apical
positions in Na-CCOC. A comparison of the superfluid density
$\rho_s\propto\sigma_{sc}(0)$ of La214 and Na-CCOC at the same level
of doping reveals that in the latter one $\rho_s$ is reduced by more
than a factor of two. Note that a pronounced difference between the
superfluid density of optimally doped La214 and Na-CCOC was also
mentioned by Kim {\it et al.}\cite{Kim06} They conclude that the
reduction of $\rho_s$ in Na-CCOC is due to a decrease of the charge
carrier concentration $n_s$. We suggest the following reason for a
possible decrease of $n_s$: The phase diagram of cuprates is usually
interpreted in terms of holes doped into the planar
Cu$d_{x^2-y^2}$-O$p_\alpha$ ($\alpha=x,y$) antibonding band. In
La$_{2-x}$Sr$_x$CuO$_4$ it is assumed that one hole per Sr atom
enters this band. However, recent $ab-initio$ calculations yielded
additional features appearing on doping of
La$_{2-x}$Sr$_x$CuO$_4$.\cite{Perry02} According to these
calculations part of the holes occupy the Cu$d_{3z^2-r^2}$-O$p_z$
orbitals. Experimentally, the existence of O $2p$ holes on the $p_z$
orbitals of the apical oxygen  were observed for La214 by
polarization-dependent fluorescence yield absorption measurements,
\cite{Chen92} and further supported by neutron diffraction
studies.\cite{Bozin03} In addition, two superconducting condensates
with $d-$ and $s-$wave symmetries were recently observed in slightly
overdoped La$_{1.83}$Sr$_{0.17}$CuO$_4$.\cite{Khasanov07_La214} It
was suggested that the $s-$wave contribution to the total superfluid
density arises from the out-of plane band related with the
Cu$d_{3z^2-r^2}$-O$p_z$ orbitals.\cite{Khasanov07_La214} Bearing in
mind that for each particular HTS family the transition temperature
$T_c$ is determined by the number of holes in the CuO$_2$ planes,
the smaller superfluid density in Na-CCOC in comparison with that in
La214 can naturally be explained by a substantial difference in the
amount of apical holes in these compounds.

\subsection{The 2D superfluid density of Na-CCOC} \label{sec:discussions_2D-superfluid}

\begin{figure}[htb]
\includegraphics[width=0.9\linewidth]{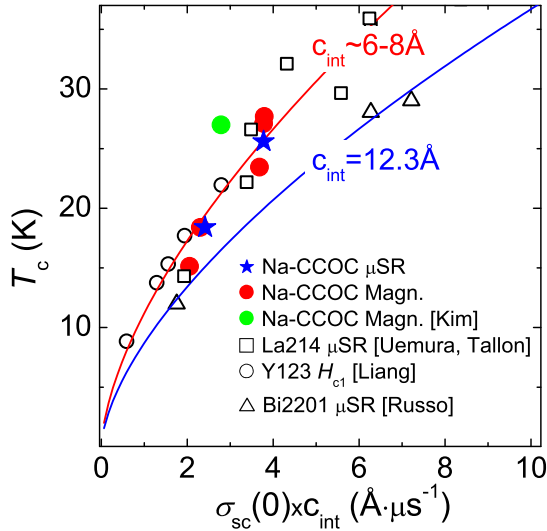}
%
\caption{(Color online) Dependence of the transition temperature
$T_c$ on the 2D superfluid density
$n_{s2D}/m^\ast=\sigma_{sc}(0)\times c_{int}$ for Na-CCOC and
various hole-doped HTS's with oxygen on the apical site. Solid red
circles and blue stars are Na-CCOC data obtained in the present
study. The green circle is the data point for Na-CCOC ($x=0.18$)
from Ref.~\onlinecite{Kim06}. Open squares are La214 data from
Refs.~\onlinecite{Uemura91} and \onlinecite{Tallon03}. Open
triangles are Bi2201 data taken from Ref.~\onlinecite{Russo07} and
open circles are Y123 data from Ref.~\onlinecite{Liang05}. The solid
lines correspond to the power law with $\sigma_{sc}(0)\times
c_{int}\propto T_c^{1.6}$. }
 \label{fig:uemura-2D}
\end{figure}

In order to check further the 2D nature of Na-CCOC,  we plot in
Fig.~\ref{fig:uemura-2D} the transition temperature $T_c$ as a
function of the 2D superfluid density, obtained as
$n_{s2D}/m^\ast=\sigma_{sc}(0)\times c_{int}$, for the same HTS's as
presented in Fig.~\ref{fig:uemura} ($c_{int}$ is the distance
between the superconducting CuO$_2$ planes: $c_{int}=7.56$~${\rm
\AA}$ for Na-CCOC,\cite{Zhigadlo07} 12.3~${\rm \AA}$ for
Bi2201,\cite{Russo07} and $\simeq6-7$~${\rm \AA}$ for Y123 and
La214\cite{Russo07}). As pointed out in Refs.~\onlinecite{Russo07}
and \onlinecite{Uemura97}, HTS's with a small $c_{int}$ tend to have
a high $T_c$ for a given $n_{s2D}/m^\ast$, following the relation
$T_c\propto 1/c_{int}$. Recalling the close similarities between
Na-CCOC and Bi2201, one expects that data points for these two
systems would exhibit the same trend (similar to what is observed in
Fig.~\ref{fig:uemura}). It is seen, however, that the data points
for Na-CCOC, Y123, and La214 (except the data points for optimally
doped and overdoped La214 that are not shown) almost follow the same
curve, while those for Bi2201 exhibit a higher 2D superfluid
density. This implies that substitution of Cl on the apical site not
only leads to a pronounced 2D-like behavior due to a reduction of
the coupling between the superconducting CuO$_2$ planes, but also to
a decrease of the transition temperature $T_c$.

\section{Conclusions}\label{sec:conclusion}

Ca$_{2-x}$Na$_x$CuO$_2$Cl$_2$ is a structural analogue to the
cuprate superconductor La$_{2-x}$Sr$_x$CuO$_4$ with Cl atoms
replacing oxygen on the apical sites. In order to check the role
of the apical oxygen for high-temperature superconductivity, we
performed $\mu$SR and magnetization studies of the in-plane
magnetic penetration depth $\lambda_{ab}$ for
Ca$_{2-x}$Na$_x$CuO$_2$Cl$_2$ samples with $x\simeq0.11$, 0.12,
0.15, 0.18, and 0.19. The following results were obtained:
The absolute value of the in-plane magnetic penetration depth at
$T=0$ was found to increase with decreasing doping from
$\lambda_{ab}(0)$=316(19)~nm for the $x\simeq0.19$ sample to
$\lambda_{ab}(0)$=430(26)~nm for the $x\simeq0.11$ one.
Comparison of the superfluid density
$\rho_s\propto\lambda_{ab}^{-2}(0)\propto\sigma_{sc}(0)$ of Na-CCOC
with that for the structurally related La214 compound reveals that
for the same doping level $\rho_s$ in Na-CCOC are is more than a
factor of two smaller than in La214. The reason for this is very
likely due to a substantial decrease of the amount of holes on the
apical sites in Na-CCOC.
Based on a comparison of the three-dimensional
[$\rho_s\propto\sigma_{sc}(0)$] and the two-dimensional
[$n_{s2D}/m^\ast=\sigma_{sc}(0)\times c_{int}$] superfluid density
of Na-CCOC with that of Bi2201 it is concluded that replacing apical
oxygen by chlorine, first, decreases the coupling between the
superconducting CuO$_2$ planes and, second, leads to a substuntial
reduction of the transition temperature $T_c$.
In addition, the appearance and enhancement of magnetism with
increasing magnetic field was observed for the optimally doped
Ca$_{1.82}$Na$_{0.18}$CuO$_2$Cl$_2$ sample, suggesting that even
optimally doped Na-CCOC has a competing magnetic state very close in
free energy to its superconducting state.
In conclusion, substitution of apical oxygen by clorine strongly
affects the superconducting and the magnetic properties of the
cuprate superconductor Ca$_{2-x}$Na$_x$CuO$_2$Cl$_2$.

\section{Acknowledgments}

This work was partly performed at the Swiss Muon Source (S$\mu$S),
Paul Scherrer Institute (PSI, Switzerland). The authors are
grateful to A.~Amato and R.~Scheuermann for assistance during the
$\mu$SR measurements and S.~Weyeneth for the help to prepare the
manuscript. This work was supported by the Swiss National Science
Foundation, the K.~Alex~M\"uller Foundation, and in part by the
NCCR program MaNEP.

\end{document}